\documentstyle[preprint,revtex]{aps}
\begin{document}
\begin{title}
{\Large Theory of p-wave Pairing for $^{3}$He on Grafoil} \\
\end{title}
\bigskip
\bigskip
\author{ {\large Andrey V. Chubukov}}
\begin{instit} \normalsize
Department of Physics, Yale University, New Haven, CT 06511-8167 \\
and \\
P.L. Kapitza Institute for Physical Problems, Moscow, Russia
\end{instit}
\bigskip
\bigskip
\author{{\large Alexander Sokol}}
\begin{instit} \normalsize
Department of Physics,
University of Illinois at Urbana-Champaign, \\
1110 West Green St., Urbana, IL 61801-3080 \\
and \\
L.D.~Landau Institute for Theoretical Physics,
Moscow, Russia
\end{instit}
\bigskip
\bigskip
\begin{abstract}
The specific heat and susceptibility data for $^3$He on
Grafoil are analyzed in the framework
of the Landau Fermi liquid theory.
The dominant interaction between
$^3$He quasiparticles is found to be in
the p-wave channel for most experimentally accessible areal
densities of
$^{3}$He. This interaction is attractive and gives rise to the p-wave
transition temperature which for moderate areal densities
is estimated to be
on the scale of several millikelvin.
The relevance of these results to the anomaly
in the specific heat observed at $ T_k = 3.2 \, \mbox{mK} $
is discussed.
\end{abstract}
\pacs{PACS: 67.50-b, 67.70+n, 67.50Dg}
\narrowtext

Experiments on monolayer films of $^3$He absorbed on Grafoil
\cite{Greywall,Greywall:Busch,Lusher}
and on  $^3$He-$^4$He mixture films on Nucleopore substrate
\cite{Higley,Bhattacharyya}
provide an opportunity to study in detail the properties of
2D Fermi liquids. Interest in
this subject has grown since the validity of the conventional Fermi
liquid theory in 2D was questioned in the context of the high-$T_c$
superconductivity \cite{Anderson}.
In the present approach, we focus on the 2D Fermi liquid at
sufficiently
low densities where perturbative calculations \cite{Engelbrecht:1}
do not show any divergencies
which might signal the breakdown of the Fermi liquid description.
Accordingly, we assume that the Fermi liquid picture is valid for
$^3$He in two dimensions.

The most remarkable property of bulk $^3$He is a superfluidity
with nonzero angular momentum $ l \! = \! 1 $ \cite{Leggett}.
However, in ``surface'' $^3$He the superfluid transition has not
yet been
observed. In what follows, we analyze the possibility of
superfluidity for $^3$He on Grafoil on
the basis of information which can be inferred
from experiments on specific heat \cite{Greywall} and magnetic
susceptibility \cite{Lusher}.
We find that for most experimentally accessible densities,
this system cannot be described by a momentum independent
(i.e., s-wave) interaction. Instead,
the dominant interaction component is in the p-wave channel.
The corresponding scattering amplitude is attractive
so that for moderate densities
one might expect to get relatively high $ T_c \sim 100$mK.
However, our calculation yields an anomalously small prefactor
in the expression for $ T_c $ in 2D which shifts the transition
down to the millikelvin region. The calculated value of
$T_c$ is reasonably close to
$3.2 \, \mbox{mK}$, where the specific heat anomaly
in $^3$He on Grafoil has been observed \cite{Greywall}.
We argue that this anomaly may correspond to the onset of
superfluidity.

We start with a brief review of the properties of dilute Fermi
liquids \cite{StatPhys}. At small densities, the s-wave component
of the scattering amplitude
is dominant. The perturbative expansion holds in powers
of the product
of the scattering amplitude and the density of states
at the Fermi surface, $ N_0 $.
In three dimensions
$ N_0 \propto p_F $ while the s-wave scattering amplitude
(scattering length)
is normally
of the same order as the range of the interaction potential, $ r_0 $.
In two dimensions, the density of states
$ N_0 $ is independent of $\rho $, but a low density expansion is
still possible
because the scattering amplitude in 2D
tends to zero logarithmically as $ \rho \to 0 $ \cite{Bloom:Bruch}.
Accordingly, the expansion parameter for the 2D problem is
$ g(\rho) = 1/\log(\rho_0/\rho) $, where
$\rho_0 \sim r_0^{-2} $.  When $ \rho \ll \rho_0 $,
the expressions for the effective mass $ m^* $ and
magnetic susceptibility $ \chi $ in 2D are \cite{Engelbrecht:2}:
$$
\frac{m^*}{m_3} = 1 + F^{s}_{1} \simeq  1 + 2g^2,
$$
\begin{equation}
\frac{\chi}{\chi_3} = \frac{1 + F^s_1}{1 + F^a_0}
\simeq  \frac{m^*}{m_3} \ \left( 1 - 2g + 4g^2\ln2 \right)^{-1},
\label{swaveonly}
\end{equation}
where $ m_3 $ and $ \chi_3 $ refer to an ideal gas of
$^3$He atoms.

The experimental data on the density dependence of the
specific heat and magnetic susceptibility are presently
available for $^3$He on Grafoil
\cite{Greywall,Greywall:Busch,Lusher} and for $^3$He-$^4$He films on
the Nucleopore substrate \cite{Higley,Bhattacharyya}.
The latter case is more difficult to analyse because
a $^3$He atom occupies a surface bound state on top of
$^{4}$He and its  hydrodynamic mass is substantially
larger than the atomic mass of $^3$He
due to the interaction with underlying $^4$He layers
\cite{Andreev:Edwards}.
In what follows, we concentrate solely on the properties
of the $^3$He film on Grafoil.
The experimental results for the effective mass \cite{Greywall}
along with theoretical predictions for
the s-wave and p-wave scattering amplitudes
are presented in Fig.\ref{fig1}.
Although the higher order terms in $g$ may be important
for larger densities, it is clear from Fig.\ref{fig1} that
the (logarithmic) density dependence of $g$ is
too weak to account for the fast increase of the effective mass
as density increases.
The above discrepancy
signals that in the experimentally accessible region of densities
the $l \! = \! 0$ harmonic does not overshadow
the higher angular momentum components.
In general, in this situation all harmonics should
have equal strength. However, we found that the plot of
$ m^*/m $ versus $ \rho $ is well described by a
simple fit which involves {\em only}
the p-wave component of the scattering
amplitude (Fig.\ref{fig1}):
\begin{equation}
\frac{m^*}{m} \approx
\left( 1 - \frac{\rho}{0.062 \, \mbox{atoms/\AA$^2$}} \right)^{-1}.
\label{pwavefit}
\end{equation}
The nearly linear behavior of $ m/m^* $ as a function
of $ \rho $ which is seen in the experiments
\cite{Greywall} even at relatively
small $ \rho $ indicates that for $^3$He absorbed on Grafoil
the p-wave
component of the scattering amplitude is anomalously large so that it
overshadows the contribution from the s-wave channel
in nearly the entire experimentally accessible range of densities.
This fact inspires us to reexamine the low-energy
expansion for $^{3}$He on Grafoil.

To proceed with the theoretical description,
consider first the case when the Born approximation is valid, i.e.
the scattering amplitude coincides with the spin independent
interaction potential $ U({\bf k}' - {\bf k}) $. Simple algebra
then yields
$ F^a = - (m^*/2\pi \hbar^2) U({\bf k}'-{\bf k}) $,
$ F^s = (m^*/2\pi \hbar^2) ( 2 U(0) - U({\bf k}'-{\bf k}) ) $.
Here both $ {\bf k}'$ and $ {\bf k} $ are on the Fermi surface,
so that $ F^{a,s} $ depend only on the angle $ \theta $ between
them.
The interaction potential can be expanded at low densities as
$ U({\bf k}'-{\bf k}) = U(0) + \lambda ({\bf k}'-{\bf k})^2 +...=
U(0) + 2 \lambda p_F^2 (1 - \cos \theta ) + O(p^{4}_{F})$.
This immediately yields
the following expressions for the effective mass (same as
(\ref{pwavefit})) and spin susceptibility:
\begin{equation}
\frac{m_3}{m^*} \simeq 1 \! - \! \lambda m_3 \rho, \
\frac{\chi_3}{\chi} \simeq \frac{m_3}{m^*}
\left( 1 \! - \! 2g \! - \! 2\lambda m^{*} \rho \right).
\label{Born}
\end{equation}

Leaving the detailed comparison with the experiment to the
study of the effects beyond the Born approximation, we merely
conclude at this point that in order to account for the increase
of the effective mass, $ \lambda $ should be positive so that
the pairing interaction in the p-wave channel
is attractive \cite{Chubukov}.
The coupling constant for the p-wave pairing is of the order of
$ 1 - m_3/m^* $.
Since the experiment shows that the effective mass may well exceed
the bare mass and $ \epsilon_F $ is nearly $1$K,
it is not clear {\em a priori} why no superfluid transition has been
observed down to the millikelvin temperature region.
To address this issue, we now
calculate explicitly the transition temperature in a
2D Fermi gas with p-wave attractive potential and show that the
prefactor in $ T_c $ is anomalously small in two dimensions.

A way to calculate the prefactor in $T_{c}$ in a weak coupling
approximation is to start with perturbation theory and collect all
second order contributions to the pairing vertex
which come from the integration over momenta far from the Fermi
surface \cite{Gorkov}. The renormalized vertex should
then be substituted
into the Cooper channel and the integration within the ladder should
be restricted to a region close to the Fermi surface where the
logarithmic term is dominant.
By carrying out the above procedure, one obtains
an equation for $T_{c}$ \cite{Gorkov}:
\begin{equation}
1 = g (1 + \alpha g) \ln \frac{\epsilon_F}{T_c} + O(g^2),
\label{Tceq}
\end{equation}
where $ g $ is a coupling constant and $ \alpha $ is a numerical
factor. Solving (\ref{Tceq}),
one gets $ T_c = \bar{\epsilon}_F \exp(-1/g) $, where
$\bar{\epsilon}_F = \epsilon_F \exp(\alpha)$.

Although for any interaction strength
the weak coupling approximation is valid
at sufficiently low density, the renormalized p-wave
vertex in the Cooper channel
$ \Gamma ({\bf p},-{\bf p};{\bf p}',-{\bf p}')
\equiv \Gamma(\theta) $ ($\cos \theta = {\bf pp}'/p_F^2 $) generally
cannot be expressed in terms of a single parameter
because the total p-wave scattering amplitude
is an unknown nonuniversal quadratic
function of the momenta. In other words, unlike for the
s-wave case, here
one can not substitute the same scattering amplitude
into all vertex functions
in the second order diagrams which contribute to $ \Gamma(\theta) $.
In view of this, we first  perform the calculation assuming
that the Born approximation is valid.
Then we reconsider
the problem by taking into account some of the effects
beyond the Born approximation.

There are four second order
diagrams which contribute to $ \Gamma (\theta ) $.
Three of them are from the
zero sound channel (Fig.\ref{fig2}a-c) while the
fourth is from the Cooper channel (Fig.\ref{fig2}d).
The evaluation of the zero sound diagrams is lengthy but
straightforward. We take advantage of the fact that
in the region of experimental interest
the s-wave component of the interaction
is considerably smaller than the p-wave component and
neglect $ U(0) $, i.e. substitute
$ U( {\bf k}_1 \alpha  {\bf k}_2 \beta ; {\bf k}_3
\gamma {\bf k}_4 \sigma )
=  \lambda ({\bf k}_1  - {\bf k}_3)^2
\, \delta_{\alpha\gamma} \delta_{\beta\sigma} $
into the vertices of Fig.\ref{fig2}.
The calculation then yields
\begin{equation}
\Gamma_{a-c} = \frac{97m \lambda^2 p_F^4 }{15\pi \hbar^2 }
\,\cos\theta
\left( \delta_{\alpha\gamma} \delta_{\beta\sigma} +
\delta_{\alpha\sigma} \delta_{\beta\gamma} \right).
\label{ac}
\end{equation}
The last diagram (Fig.\ref{fig2}d) contains both
$ \log ( \epsilon_F/T_c ) $ and
the contribution to the prefactor, and also contributes
to the vacuum renormalization
which transforms the interaction potential
into the scattering amplitude
\cite{StatPhys}.
The vacuum renormalization has to be subtracted from
the diagram of Fig.\ref{fig2}d. This procedure eliminates
ultraviolet divergence in the theory leaving $ \epsilon_F $
as the only dimensional quantity:
\begin{equation}
\Gamma_d = \! - \frac{f_1^2 m \hbar^2 }{4\pi }
\ln \! \left[ \frac{2 \epsilon_F \gamma}{\pi e T} \right] \!
\cos\theta
\left( \delta_{\alpha\gamma} \delta_{\beta\sigma} +
\delta_{\alpha\sigma} \delta_{\beta\gamma} \right),
\label{d}
\end{equation}
where $ \ln\gamma $ is the Euler constant $ C \approx 0.577 $ and
$ f_1 $ is the p-wave scattering amplitude, which in the Born
approximation is equal to $ - 2 \lambda p_F^2 / \hbar^2 $.

Expressing (\ref{ac}) in terms of $ f_1 $ and
combining it with (\ref{d}),
we get the renormalized pairing interaction in the p-wave channel
$ \Gamma = \Gamma_{a-c} + \Gamma_d $
which includes both leading (logarithmic) and next to
leading order terms in $ f_1 $.
The instability criterion then yields ($ f_1 < 0 $):
\begin{equation}
T_c = \frac{ 2\gamma \epsilon_F}{\pi }
\ \exp \left( -\frac{112}{15} \right)
\, \exp \left( -\frac{4\pi}{m|f_1|} \right).
\label{Tc}
\end{equation}
While $ 2\gamma/\pi \approx 1.13 $,
the factor
$ \exp ( -112/15 ) \approx 5.72 \cdot 10^{-4} $
reduces $T_c$ in (\ref{Tc}) by
more than three orders of magnitude.
As a result, even if the coupling constant
$ g_p = m|f_1|/4\pi \sim 1$, the critical temperature
$T_{c} \sim 10^{-4} \epsilon_F $.

The results above were obtained in the Born approximation,
i.e. under the
assumption that the
Born parameter $u \simeq \lambda m_{3} /4\pi r_0^2  \ll 1$.
The validity of the Born approximation
is, however, questionable for $^{3}$He on Grafoil. Indeed,
it follows from (\ref{Born}) that in the Born approximation
the linear in density term in $ F_0^{a} $
is two times larger than that in $ F_1^{s} $.
However, the experimental results \cite{Greywall,Lusher}
give a much smaller value for this ratio.
This suggests that the Born parameter is, in fact, not small
and hence the density-independent
(``vacuum'') corrections to the Fermi liquid
parameters are important. We calculated the leading vacuum
p-channel corrections in the symmetrical gauge and obtained
\begin{eqnarray}
F_1^s & = & \lambda m^{*} \rho
\left( 1 + \frac{9u}{4} \right), \nonumber \\
F_0^a & = & -2g -2\lambda m^{*} \rho \left( 1 - \frac{u}{4} \right).
\label{vacuum}
\end{eqnarray}
Since $ \lambda $ should be positive in order to account
for the increase
of $ m^*/m_3 $ with the density, vacuum corrections increase
the linear
term in $ F_1^s $ and reduce that in $ F_0^a $ which narrows the gap
between theory and experiment.
A fit to the experiment gives $ u \sim 1\! -\! 2 $, that is the
vacuum corrections are, indeed, strong (the uncertainty in $u$ is
related to the s-wave contribution to $F_0^a$ which is difficult
to estimate precisely).

The next step would be to calculate $T_{c}$ beyond the Born
approximation.
However, we already mentioned that in order
to solve the problem one needs to
know what the renormalized scattering amplitudes are
for all combinations of the momenta
relevant to the diagrams of Fig.\ref{fig2}.
Such calculation is lengthy and not very
informative because it is not clear whether one can deal
with only the
leading term in $u$ in the region of parameters
relevant to the experiments.
Because of this complication, below we
use a more qualitative approach and just take into account  the fact
that the spin structure of
the p-wave interaction potential does not survive
the effect of vacuum renormalization, i.e. the total p-wave
scattering amplitude
has both spin-independent and spin-dependent parts even
if the initial
interaction was spin-independent.  Accordingly, we model the effect
of vacuum renormalization by introducing an effective
potential which satisfies the Born approximation condition and
has both spin-independent and spin-dependent parts
\cite{pseudopotential}:
\begin{equation}
U_{eff} = U({\bf k}' - {\bf k}) \delta_{\alpha\gamma}
\delta_{\beta\sigma}
+ \bar{U}({\bf k}' - {\bf k})
\vec \sigma_{\alpha\gamma} \vec \sigma_{\beta\sigma}.
\label{Ubar}
\end{equation}
The low density expansion of $ \bar{U} $ is
$ \bar{U}(\theta)
= \bar{U}(0) + 2 \bar{\lambda} p_F^2 (1 - \cos \theta) + O(p_F^4) $.
This effective potential reproduces measured $ m^*(\rho) $ and
$ \chi(\rho) $ quite well for
$ \zeta = \bar{\lambda}/(\bar{\lambda} + \lambda) \approx 0.3-0.4 $
(note much smaller uncertainty in $\zeta$ than in $u$) .

The calculation of $ T_c $ with $ U_{eff} $
proceeds along the same lines
as above.
We  skip the intermediate calculations and present
only the result:
\begin{equation}
T_c = \frac{ 2\gamma \epsilon_F}{\pi }
\exp \left( -\frac{112}{15} + \frac{56}{3} \, \zeta
- \frac{308}{15} \, \zeta^2 \right)
\exp \left( -\frac{1}{g_p} \right).
\label{Tceff}
\end{equation}
The p-wave coupling constant now is
\begin{equation}
g_{p} = \frac{mp^{2}_{F}}{2\pi \hbar^2 } (\lambda + \bar{\lambda}).
\end{equation}
As it turns out,
$ \zeta = 0.3\! -\! 0.4 $, inferred from the fit to the experiment,
corresponds to the
prefactor $ \bar{\epsilon}_F = 0.03\! -\! 0.04 \,
\epsilon_{F} $. This value is larger than our earlier estimate though
is still considerably smaller than
the p-wave result in three dimensions,
$ 0.1 \epsilon_{F}$ \cite{Chubukov:Kagan}.
It is worth mentioning here
that for bulk $^3$He the value of the prefactor
inferred from the measured $T_c$
was found to be rather insensitive to the
particular form of the interaction \cite{Bedell:Pines}.

In order to estimate $T_{c}$,
we express the coupling constant $g_p$ as
\begin{equation}
g_{p} = \frac{1 - m_3/m^*}{1 + 2 \zeta}.
\end{equation}
For small densities, $m^{*} \approx m_{3}$ and $T_{c}$ is
exponentially small. However, for moderate densities, $m^{*}$ is
significantly larger
than the bare mass and the coupling constant saturates at the value
$(1 + 2\zeta)^{-1} \approx 0.5\! -\! 0.6 $.
Substituting this value of $ g_p $
into (\ref{Tceff}), we get $ T_c \sim 5\! -\! 7 \, \mbox{mK} $
which is of the same order
as $ T_k \approx 3.2 \, \mbox{mK} $
where the anomaly in the specific heat
has been observed.
Although the quantitative agreement with the experiment
is unanticipated because of the approximate nature
of the theoretical considerations, our
results indicate that the dense
``surface'' $^{3}$He on Grafoil may  become
superfluid in the experimentally accessible temperature range.
Note that there is no true off-diagonal long range
order in two dimensions, but the
superfluid density is finite below
the transition \cite{K-T,Fisher}. The actual
Kosterlitz-Thouless
transition  temperature does not differ
substantially from the ``mean-field'' $T_c$
which we have calculated \cite{Fisher}.

It is argued in \cite{Greywall:Busch} that
the specific heat anomaly in $^3$He on Grafoil at
$ T_k $
is an intrinsic property of the fluid monolayer.
The calculated value of $T_c$ (\ref{Tceff})
is comparable with $T_k$, but, unlike $T_k$,
it significantly drops down at low densities. We suggest
that the density
independence of $T_{k}$ observed in the experiments
may in fact be a result
of a phase separation which accompanies the superfluid transition,
that is for
arbitrary density of $^{3}$He there is an energetically stable
superfluid dense phase of $^{3}$He below $T_{k}$.
This tentative scenario explains both the lack of
density dependence in $ T_k $ and a rapid decrease
of the specific heat below $ T_k $.

To summarize, in this paper we have analysed
the specific heat and susceptibility
data for $^{3}$He on Grafoil.
We found that the p-wave component of the
interaction between
$^{3}$He quasiparticles is dominant
for all experimentally
accessible areal densities of $^{3}$He.
This interaction is
attractive and gives rise
to a p-wave transition temperature which for the dense
``surface'' $^3$He is estimated
to be in the millikelvin temperature region.
We suggest that $^3$He on Grafoil
may be superfluid below $ 3.2 \, \mbox{mK} $ where the
anomaly in the specific heat has been observed.

We would like to thank L. Bruch, I. Engelbrecht, D. Frenkel,
M. Gelfand,
D.S. Greywall, C.J. Pethick, D. Pines, and M. Webb
for  useful discussions. This work has been supported
by NSF Grants DMR88-57228 and DMR88-17613.

\figure{The density dependence of the effective mass
for $^3$He on Grafoil.
The experimental data \cite{Greywall}
is fitted using the interaction potential with
dominant s-wave or p-wave amplitudes
(dotted and dashed lines, respectively).
\label{fig1}}

\figure{Second order
diagrams which contribute to the pairing vertex.
The wavy line represents the interaction potential. \label{fig2}}

\end{document}